\documentclass[12pt,aps,amsmath,latexsym,amsfonts]{JHEP3}

\usepackage{epsfig}
\usepackage{amsfonts,amssymb,amsmath}
\usepackage[hang]{subfigure}
\usepackage{epstopdf}

\newcommand{\be}{\begin{equation}}
\newcommand{\ee}{\end{equation}}
\newcommand{\bea}{\begin{eqnarray}}
\newcommand{\eea}{\end{eqnarray}}
\newcommand{\beal}{\begin{aligned}}
\newcommand{\eeal}{\end{aligned}}

\title{Time dependent black holes and scalar hair}

\author{
Sarah Chadburn$^1$\thanks{Email: s.e.chadburn@durham.ac.uk},
Ruth Gregory$^{1,2}$\thanks{Email: r.a.w.gregory@durham.ac.uk} \\ 
$^1${\it Centre for Particle Theory, South Road, Durham, DH1 3LE, UK}\\
$^2${\it Perimeter Institute, 31 Caroline Street North, Waterloo, ON, N2L 2Y5,
Canada}
}

\abstract{
We show how to correctly account for scalar accretion onto black holes
in scalar field models of dark energy by a consistent expansion in
terms of a slow roll parameter. At leading order, we find an analytic 
solution for the scalar field within our Hubble volume which is regular 
on both black hole and cosmological event horizons, and compute 
the back reaction of the scalar on the black hole, calculating the 
resulting expansion of the black hole. Our results are independent 
of the relative size of black hole and cosmological event horizons. 
We comment on the implications for more general black hole 
accretion, and the no hair theorems.
}

\preprint{DCPT-13/15}

\begin{document}
\newcommand{\zed}{$\mathbb{Z}_2$}

\section{Introduction}

Black holes are one of the most fascinating objects in classical
general relativity. They represent the endpoint of gravitational
collapse of a large mass object, independent of initial conditions.
The prototypical black hole, the Schwarzschild solution, was first
presented barely a year after Einstein put forward his new theory
of gravity, yet it took half a century before relativists were
confident of the interpretation of this solution, and began to construct
a rigorous set of theorems describing the properties of black holes.
One of the more poetically labelled set -- the ``no hair'' theorems 
\cite{nohair} -- has been perhaps the most contentious. Originally 
demonstrated for static vacuum black holes, \cite{Cart,Be72}, the 
no hair theorems have been refined and extended to a wide range 
of interacting particle and gravitational theories, \cite{NHF,NHG,MB}. 
Nonetheless, there are many known ``violations'' of the
no hair theorems, from the dressing of black holes by scalar or gauge
condensates \cite{hair}, to literal hair, in the form of topological
defects which extend out from the black hole to infinity, \cite{AGK}.

The issue of the most basic hair however, scalar hair, is thought to be
understood, yet is perhaps the most perplexing. For a wide range of 
potentials, it is easy to show that a static black hole cannot have
a nontrivial scalar field on its event horizon (unless the space-time is
asymptotically anti-de Sitter, \cite{TMN}, or the scalar potential has
been specifically tailored, \cite{Z}). However, within cosmology,
scalar fields are widely used, not only for inflation, but as an 
expedient model of quintessence, \cite{CDS}, or late time acceleration 
(see \cite{CST} for a review of dark energy models). 
Indeed, the standard exponential
scalar potential - widely used to give power law acceleration, was shown
to be in conflict with spherically symmetric asymptotically flat or de
Sitter static black hole space-times, \cite{PW}.
How then can these cosmological scalars interact with a 
black hole event horizon? Generally, the scalar field is varying on a very
slow time scale relative to the light crossing time of the black hole,
thus we might expect that it has little effect on the event horizon itself,
yet, like the inexorable flow of time, the scalar must evolve
cosmologically, and it seems contradictory that it is pinned to a single
value for all time at the black hole -- an issue explored by Barrow,
\cite{Barrow}, in the context of primordial black holes.

In fact, the resolution of the conflict is quite straightforward: 
The no hair theorems were explored in the context of static black
holes, and quite clearly a rolling cosmological scalar can never be 
static.  Correspondingly, the presence of a black hole in a cosmology 
implies that the cosmology cannot be spatially homogeneous, and 
therefore does not have the canonical FRW form. While we can use
cosmological perturbation theory to describe the far field effect
of the black hole, we cannot describe the full spatial geometry 
including the event horizon within the context of perturbation
theory -- and it is the effect of the event horizon which is 
critical in determining the behaviour of the scalar.
We therefore need a time-dependent black hole solution.

Exact black hole solutions are surprisingly thin on the ground,
given how ubiquitous black hole are in nature, and time dependent
ones even more so. The first dynamical black hole `solution' was
postulated by McVittie, \cite{McV}, who modified the Schwarzschild
solution to include an FRW-like cosmology, with the black hole
event horizon at fixed comoving radius and a fluid energy density 
dependent only on time.  However, as pointed out in \cite{KKM},
one expects the black hole to break the spatial homogeneity of
the cosmology, and thus (except in the special case of a de Sitter
universe) this unrealistic set-up leads to a singularity on the
would be event horizon. The McVittie solution has been generalised,
\cite{Sultana:2005tp,Faraoni:2007es}, to allow
for radial accretion, although the generalisations follow the
method of `metric engineering', in that a specific ansatz is 
proposed, then the matter stress energy inferred (see also 
\cite{Carrera:2009ve} for a discussion of some possible issues
with these approaches). 

Instead of imposing a particular metric ansatz, a successful 
approach has been instead to postulate a particular format 
for the behaviour of the solution, notably self-similarity, first
used to estimate primordial black hole accretion in \cite{CH}, 
but also extended to allow more general perfect fluid
equations of state, \cite{LCF,BH,MHC}, and scalar tensor 
gravity, \cite{Harada}. Although these solutions do make an
assumption about the behaviour of the metric, and are
restricted to a perfect fluid, they have the advantage that
they are exact, allow for time-dependence of the black hole,
to be explored without resorting to complex numerical methods.
(See \cite{Crev} for a brief review.)

There is also the time dependent Vaidya solution, \cite{Vaidya}, 
which is somewhat special, as the mass of the black
hole now becomes dependent on a null coordinate, requiring 
a null matter source. 
There are also exact solutions for Einstein plus scalar field 
which either represent a collapsing scalar field, \cite{Husain:1994uj},
or use solution generating techniques to add scalar profiles to a
black hole, \cite{Fonarev:1994xq}. In both cases however, the
would be event horizon is singular, and indeed in \cite{Husain:1994uj},
identifying event horizons becomes an issue. 
Thus a direct approach to finding an exact solution seems to have been
unsuccessful in the sense that in cosmology, we expect a solution which 
will be locally interpretable as a ``black hole'' (Schwarzschild solution)
but that nonetheless will also have large scale cosmological evolution,
which on long time scales will give a natural evolution of
the black hole event horizon area due to scalar accretion. 
(See \cite{Carrera:2008pi} for a review and discussion of the 
interplay between cosmological expansion and local systems). 

Another approach in the literature is to find a probe solution
for the scalar field, i.e.\ one in which the scalar field evolves
on a Schwarzschild background, \cite{TJ,FK}. Here, near the 
event horizon, the scalar field must be
a function of the advanced null coordinate, $v \sim e^{t+r^*}$,
which parametrises the black hole future event horizon.
One can then estimate the back-reaction on the event horizon 
using the energy momentum of this approximate solution. 
A problem with this approach however, is that the coordinates being
used are local static Schwarzschild coordinates, which do not 
correspond to the cosmological flow of time at large distances 
and therefore give no guarantee that any `solution' will be well 
behaved at cosmological event horizons.
One can also worry that estimates using local intuitive
notions of energy, rather than an analysis of the Einstein equations
with a scalar source, might be misleading. 

In order to be confident that probe calculations give a good estimate
of scalar accretion onto black holes, we require a time-dependent 
scalar field with a time dependent black hole. Here,
we present a resolution of this problem, by expanding the equations
of motion for the geometry and the scalar field order by order in
a ``slow roll'' parameter.  We present a procedure for solving
for the time dependent scalar on the black hole, finding the leading
order solution which extends from the near black hole expectation 
of \cite{TJ,FK} to the cosmological solution at large distances. 
Our solution is valid independent of the relative sizes of the black 
hole and cosmological event horizons.
We also present the back reaction of the scalar field
on the black hole, and calculate the expansion rate of the black
hole  due to scalar accretion. An advantage of our method is that
we can identify the event horizons accurately as null surfaces,
therefore the usual ambiguity of apparent vs.\ event horizons
does not occur.

It is perhaps worth emphasising that our approach is explicitly
to construct a natural extension of the Schwarzschild `vacuum' solution,
and not some engineered exact solution of Einstein's equations, either
by making a metric ansatz, or by taking an ansatz for the behaviour of
the solution. Our only input will be the symmetries of the physical set-up, 
and the output a nonsingular solution corresponding to the physical set-up 
of an FRW expanding universe at large scales, and a (Schwarzschild)
black hole with some local scalar field on small scales.

\section{Scalar fields and black holes: set-up}

In this section, we set up the equations of motion for the scalar
field with the black hole. The idea is to write down the general
equations of motion compatible with the symmetries, then to
expand them order by order in the kinetic energy of the scalar
field.

To motivate our approach, consider a universe in which the 
acceleration of the universe is driven by a slowly rolling scalar field, 
somewhat analogous to inflation though clearly at a much lower scale. 
A simple toy model consists of an FRW universe,
\be
ds^2 = dt^2 - a^2(t) d{\bf x}^2
\ee
and a scalar field with an exponential 
potential $W = M^4 e^{-\beta\phi}$,
which leads to a power law acceleration
\be
a(t) = \left ( \frac{t}{t_0} \right)^k \;\;,\;\;\;\;\;
\phi = \phi_0 + \frac{2}{\beta} \ln \frac{t}{t_0}\;\;,\;\;\;\;\;
k = \frac{2\kappa}{\beta^2} = \frac{2 W(\phi)^2}{M_p^2 W'(\phi)^2}
=\varepsilon^{-1}
\ee
where $\kappa=8\pi G=M_p^{-2}$, and $\varepsilon$ 
is the conventional slow roll parameter introduced
here to emphasise that $k\gg1$. 

Now consider the solution in conformal time $\eta 
= \eta_0 (t/t_0)^{(\varepsilon-1)/\varepsilon}$:
\be
ds^2 = \left[ \frac{\eta_0}{\eta} \right ]^{2
+\frac{2\varepsilon}{1-\varepsilon}} 
\left [ (d\eta - d\rho)(d\eta+d\rho) - \rho^2 d\Omega_{{\rm II}}^2\right]
\label{sccos}
\ee
where $d\Omega_{{\rm II}}^2$
is the standard line element on the unit sphere, and
\be
\phi = \phi_0 - \frac{2\varepsilon}{\beta(1-\varepsilon)} 
\ln \frac{\eta}{\eta_0} \;.
\label{phicost}
\ee
Then, if we assume that $\varepsilon\ll1$, and expand
$[\eta_0/\eta]^{\frac{2\varepsilon}{1-\varepsilon}} \sim 
1 - 2\varepsilon \ln |\eta/\eta_0|$, 
this metric is perturbatively
close to the de Sitter (dS) metric:
\be
g_{ab} = g^{(DS)}_{ab} \left ( 1 - 2 \varepsilon \ln 
\left | \frac{\eta}{\eta_0} \right | + {\cal O} (\varepsilon^2)\right)
\ee
over a Hubble time interval. The dark energy universe can
therefore be expressed as a linear perturbation of a known
exact solution to Einstein's equations -- the de Sitter universe.
Looking at \eqref{phicost}, we see that $\phi = \phi_0$, plus a
correction of order ${\cal O} \left (\varepsilon^{1/2} M_p \right )$, 
and thus for small $\varepsilon$, we can express the 
cosmology as a de Sitter universe with a small correction
of order ${\cal O}(\varepsilon^{1/2})$ for the scalar, and 
${\cal O}(\varepsilon)$ for the geometry.

Now consider adding a black hole to this cosmological rolling 
scalar solution. Given that the background solution is
a perturbation of a de Sitter universe, it is reasonable to
suppose that the black hole plus scalar might be 
expressible as a perturbation of a Schwarzschild de 
Sitter space-time. Note however, that this is distinct from 
conventional cosmological perturbation theory, where one
perturbs a given spatial section and evolves forward in time.
Rather, here we seek to write the fully interacting
black hole-plus-scalar system as a perturbation 
expansion around the static solution
in the kinetic motion of the scalar. 
Just as the rolling scalar cosmology is perturbatively
close to the de Sitter manifold, we expect that the
rolling scalar plus black hole manifold should be
close to the Schwarzschild de Sitter manifold, and
just as the linear expansion above has a range
of validity, we expect that our expansion will also 
be valid only over Hubble time-scales.

The geometry of the time-dependent black hole will not 
now have constant curvature spatial slices, 
but we do expect an SO$(3)$ symmetry, corresponding to the
spherical symmetry of the black hole. We also expect
both time and radial dependence in the metric. The general
metric may therefore be written in the form \cite{BCG,CG}
\be
ds^2 = 4 e^{2\nu}B^{-1/2}dUdV - Bd\Omega_{{\rm II}}^2 .
\label{weylform}
\ee
This form of the metric elucidates both the symmetry of the
space-time (SO(3)), as well as the remaining gauge freedom (the
conformal group in the $U,V$ directions). By rewriting
the coordinates in light-cone form, it will be clear how to deal 
with the event horizons present in the anticipated 
solution (the cosmological and black hole event horizons),
as well as how to change gauge to analytically extend across
these horizons. In particular, the null coordinates allow us to
identify the actual event horizons of the solution, as opposed to
apparent horizons, as an horizon is of course
always a null surface, defined by $U$ or $V=$ constant.

Using \eqref{weylform}, the coupled Einstein-scalar 
equations for the variables $B, \nu, \phi$ are
\bea
\phi_{,UV} &=& - W_{,\phi}(\phi)B^{-1/2}e^{2\nu}
- \frac{1}{2B}\left(B_{,U}\phi_{,V}+B_{,V}\phi_{,U}\right)
\label{phiUVeq}\\
B_{,UV} &=& 2\left(\kappa W(\phi)B^{1/2} - B^{-1/2}\right)e^{2\nu}
\label{BUVeq}\\
\nu_{,UV} &=& \frac{1}{2}\left(\kappa W(\phi)B^{-1/2}
+ B^{-3/2}\right)e^{2\nu} - \frac{\kappa}{2}\phi_{,U}\phi_{,V}
\label{nuUNeq}\\
B_{,VV} &=& 2\nu_{,V} B_{,V} - \kappa B \phi_{,V}^2
\label{Vint}\\
B_{,UU} &=& 2\nu_{,U} B_{,U} - \kappa B \phi_{,U}^2
\label{Uint}
\eea
where $W(\phi)$ is a general potential, the only stipulation
being that it satisfies the slow roll condition $\varepsilon\ll1$.

For a constant scalar field (i.e.\ briefly ignoring \eqref{phiUVeq})
a generalisation of the Birkhoff theorem shows that
the Einstein equations have Schwarzschild de Sitter (SdS) 
as a general solution, \cite{BCG}. Given that we follow
a similar procedure in analysing the rolling scalar, it is
worth briefly reviewing the steps of this argument.

If $\phi$ is constant, \eqref{Vint} and \eqref{Uint} 
can be integrated directly to give 
\be
2\nu = \ln B_{,V} + \ln G'(U) = \ln B_{,U} + \ln F'(V)
\ee
where $F'$ and $G'$ represent arbitrary integration functions. 
Consistency of these expressions leads us to deduce that $B$ 
must be a function of $\left[F(V) + G(U)\right]$, and hence 
\be
e^{2\nu} = F'(V) G'(U) B'(F+G)\,.
\ee
Inserting into \eqref{BUVeq} then gives
\be
\beal
F'G' B'' &= 2F'G'B' \left ( \kappa W_0 B^{1/2} - B^{-1/2} \right)\\
\Rightarrow \quad
B' &= \frac43 \kappa W_0 B^{3/2}- 4 B^{1/2} + 8GM 
\eeal
\ee
where $8GM$ is an integration constant (suggestively labeled!),
$W_0=W(\phi_0)$, and primes denote differentiation with 
respect to the argument of the function.
However, writing $N(r)$ as the SdS potential
\be
N(r) = 1 - \frac{2GM}{r} - H^2 r^2
\ee
shows that in fact
\be
B' = - 4 \sqrt{B} N(\sqrt{B})
\ee
with $H^2 = \kappa W_0/3$, the vacuum density of the constant
scalar field. Changing coordinates to
$r = B^{1/2}$, $t=2(G(U)-F(V))$, then gives
\be
ds^2 = \frac{F'G'B'}{B^{1/2}} 4 dUdV - B d\Omega_{{\rm II}}^2
\to N(r) dt^2 - \frac{dr^2}{N(r)} - r^2 d\Omega_{{\rm II}}^2
\label{UVmet}
\ee
i.e.\ the Schwarzschild de Sitter metric in static coordinates. 
In this form, we can see explicitly that the arbitrary integration 
functions $F$ and $G$ are simply gauge degrees of freedom
of the metric \eqref{UVmet}, and in fact represent the conformal 
transformations on the $(U,V)-$plane. Since $B'$ can vanish, 
this metric will in general have singularities at certain values of
$B$. These are none other than the black hole and cosmological
event horizons of the static co-ordinates. However, ``cosmological'' 
coordinates at large `$r$' would not have an horizon, and
would asymptote a standard cosmological
de Sitter space-time; we therefore need to identify the (Kruskal) 
transformations which provide extensions across each
horizon. 

Writing $r^*$ as the usual tortoise co-ordinate, note that 
\be
r^* = \int \frac{dr}{N(r)} = - 2 \int \frac{dB}{B'} = -2(F+G)
\ee
thus $t-r^* = 4G$, $t+r^* = -4F$. Following the usual Kruskal
method, we now choose the functions $F(V)$ and $G(U)$ to
make the metric regular at the cosmological event horizon $r_c$:
\be
\beal
F(V) &= -\frac{1}{2N'(r_c)} \ln \left [ N'(r_c) V \right]\;,\;\;
&{\rm or} \;\;
V &= R_c \exp \left [ \frac{(t+r^*)}{2R_c}\right] \\
G(U) &= \frac{1}{2N'(r_c)}\ln \left [N'(r_c) U\right] \;,\;\;
&{\rm or} \;\;
U &= R_c \exp \left [ \frac{(t-r^*)}{2R_c}\right] 
\eeal
\label{kruskcos}
\ee
where we have written $R_c = 1/N'(r_c)$ as shorthand
(see appendix).

Thus, the original functions of the metric \eqref{weylform}
are
\be
\beal
B(U,V) &= r^2 = \left ( {r^*}^{-1} \left [R_c \ln \frac{V}{U}
\right ] \right)^2\\
e^{2\nu} &= F'G'B' = \frac{R_c^2\sqrt{B}N(\sqrt{B})}{ UV}
= \frac{R_c^2}{UV} r N(r)
\eeal
\ee
where ${r^*}^{-1}$ is the inverse tortoise function, which does
not in general have a closed analytic form, and $r$ is 
understood to be a function of $U$ and $V$.
In these co-ordinates, as $r\to r_c$, 
\be
V = U \exp \left [ \frac{r^*}{R_c} \right] \approx U (r-r_c)
\ee
thus the cosmological event horizon is at $V=0$ and is 
parametrized by $U$. Moreover 
\be
e^{2\nu} = - \frac{R_c^2}{U^2} \frac{r_cN}{(r-r_c)}
\ee
is explicitly regular as expected.

For future reference, the Kruskal extension at the black hole 
event horizon $r_h$ would be given by the null co-ordinate choice
\be
\beal
F(v) &= -\frac{1}{2N'(r_h)} \ln \left [ N'(r_h) v \right]\;,\;\;
&{\rm or} \;\;
v &= R_h  \exp \left [ \frac{(t+r^*)}{2R_h}\right] \\
G(u) &= -\frac{1}{2N'(r_h)}\ln \left [-N'(r_h) u\right] \;,\;\;
&{\rm or} \;\;
u &= R_h \exp \left [ -\frac{(t-r^*)}{2R_h}\right] 
\eeal
\label{kruskhor}
\ee
writing $R_h=1/N'(r_h)$ as before. The black hole event
horizon is at $u=0$, and parametrized by $v$.
We will mostly work with the `cosmological' co-ordinates
$U$ and $V$, however, we will refer to the black hole
Kruskals $(u,v)$ when checking regularity at the event 
horizon.

Now suppose that we take into account that $\phi$ is not 
constant, and write
\be
\beal
\phi &= \phi_0 + \sqrt{2\varepsilon} M_p \phi_1(U,V)\\
B &= r^2 \left ( 1 + \varepsilon \delta_1(U,V) \right)\\
\nu &= \nu_0 + \varepsilon \delta_2(U,V)
\eeal
\ee
then, recalling the expressions for $B_0=r^2$ and $\nu_0$,
and expanding the equations of motion shows that the
equation for $\phi_1$ is at order ${\cal O}(\sqrt{\varepsilon})$, 
and decouples from the perturbations to the geometry, which
appear at order ${\cal O} (\varepsilon)$
\bea
\sqrt{2\varepsilon}\,\phi_{1,UV} &=& \sqrt{2\varepsilon}\,
\frac{N}{UV} \left [ 3H^2 R_c^2 + \frac{R_c}{r}
\left (V\phi_{1,V} - U \phi_{1,U} \right ) \right]
\label{phiUVeqeps}\\
\varepsilon r^2 \delta_{1,UV} &=& - 2\varepsilon R_c \frac{rN}{UV}
\left ( U \delta_{1,U} - V\delta_{1,V} \right) \label{BUVeps}\\
&&+\varepsilon R_c^2 \frac{N}{UV} \left [ 4\delta_2(3H^2r^2-1)
+ 3 \delta_1 (1-H^2r^2) - 12 H^2 r^2 \phi_1 \right ]
\nonumber\\
\varepsilon r^2 \delta_{2,UV} &=& \varepsilon \frac{R_c^2N}{UV}
\left[ (1+3H^2 r^2) \delta_2 - \frac{3\delta_1}{4}(1+ H^2 r^2) 
-3H^2r^2\phi_1\right] - \varepsilon r^2 \phi_{1,U} \phi_{1,V}
\quad \label{nuUNeqeps}\\
\varepsilon r^2 \delta_{1,VV} &=& 
4\varepsilon R_c \frac{rN}{V} (\delta_{2,V} - \delta_{1,V})
+\varepsilon \frac{\delta_{1,V}}{V} \left ( rR_c (rN)' - r^2 \right)
- 2 \varepsilon r^2 \phi_{1,V}^2
\label{Vinteps}\\
\varepsilon r^2 \delta_{1,UU} &=& 
4\varepsilon R_c \frac{rN}{U} (\delta_{1,U} - \delta_{2,U})
- \varepsilon \frac{\delta_{1,U}}{U} \left ( rR_c (rN)' + r^2 \right)
- 2 \varepsilon r^2 \phi_{1,U}^2
\label{Uinteps}
\eea
We therefore solve first for the scalar field rolling in the 
SdS background, then compute the back-reaction on
the geometry.

\section{The scalar field}

In order to solve \eqref{phiUVeqeps}, it is most transparent
to present the equation in terms of our SdS variables:
\be
\ddot{\phi}_1 - \frac{1}{r^2} \frac{\partial~}{\partial r^*} 
\left ( r^2 \frac{\partial\phi_1}{\partial r^*}\right) = 3 H^2 N(r)
\label{phipt}
\ee
Clearly, this equation will have oscillatory solutions for $\phi_1$,
corresponding to partial waves scattering off the black hole, 
however, we are interested in the background, `vacuum'
solution where $\phi_1$ rolls according to the potential $W$.
Thus we set
\be
\phi_1 = \lambda t + \varphi(r)
\label{gendphi}
\ee
where \eqref{phipt} gives
\be
\frac{d~}{dr} \left ( r^2 N \frac{d\varphi}{dr}\right) =
-3H^2 r^2 
\ee
which is solved by
\be
\varphi = - \sum_i \left [ H^2 r_i
+ \frac{C}{ r^2_i} \right] R_i \ln | r - r_i| 
+\frac{C}{2GM} \ln r
\ee
with $C$ an integration constant. (See the appendix for 
definitions of the $R_i$ etc.~together with useful identities.)

For a nonsingular solution, the $\phi$ field must be regular 
in a locally regular coordinate system at both the black hole 
($r_h$) and cosmological ($r_c$) future event horizons. 
At the cosmological event horizon the appropriate co-ordinates
are $(U,V)$, with $V\to0$ at the cosmological event horizon, and 
$t\sim R_c \ln V$, $r^* \sim R_c \ln |r-r_c| \sim R_c\ln V$. 
Conversely, using \eqref{kruskhor} near the black hole event
horizon shows that $u\to0$, with $t\sim -R_h \ln(-u)$, and 
$r^*\sim R_h \ln(r-r_h)\sim R_h \ln(-u)$. Therefore, 
demanding regularity of $\phi_1$
gives two constraints on $\lambda$ and $C$:
\be
\lambda - H^2 r_c - \frac{C}{r^2_c} =0
=\lambda + H^2 r_h + \frac{C}{r^2_h}
\ee
solved by
\be
C = -H^2 \frac{r_h^2 r_c^2 (r_h+r_c)}{r_h^2+r_c^2}\;\;\;,\;\;\;
\lambda = \frac{(r_c-r_h)}{r_h^2+r_c^2}\;.
\ee
revisiting the expression for $\varphi$, we see 
\be
\varphi = - \lambda R_c \ln|r-r_c| + \lambda R_h \ln(r-r_h)
+ \lambda \,\frac{r_c^2R_h - r_h^2R_c}{2r_hr_c}\,\ln(r-r_N)
-\frac{r_hr_c}{r_h^2+r_c^2} \ln r
\label{varphiexpression}
\ee
(using various identities from the appendix).

Pulling this together we can write the $\phi$ field in the 
Kruskal coordinates (remembering that $r=r(V/U)$ or $r(uv)$)
\be
\begin{aligned}
\phi &= \phi_0 + \sqrt{2\varepsilon} M_p\, \lambda
\Bigl [ 2 R_c \ln \frac{U}{R_c} + 2 R_h \ln(r-r_h) 
+ \frac{r_c R_h}{r_h} \ln(r-r_N)- \frac{r_h r_c \ln r}{(r_c-r_h)}
\Bigr] \qquad \\
&= \phi_0 + \sqrt{2\varepsilon} M_p\, \lambda
\Bigl [ 2 R_h \ln \frac{v}{R_h} - 2 R_c \ln|r-r_c| 
- \frac{r_h R_c}{r_c} \ln(r-r_N)- \frac{r_h r_c \ln r}{(r_c-r_h)}
\Bigr]
\end{aligned}
\ee
which is manifestly nonsingular at the horizons, and illustrated
in figure \ref{fig:phicontours}.
\FIGURE{
\includegraphics[scale=1]{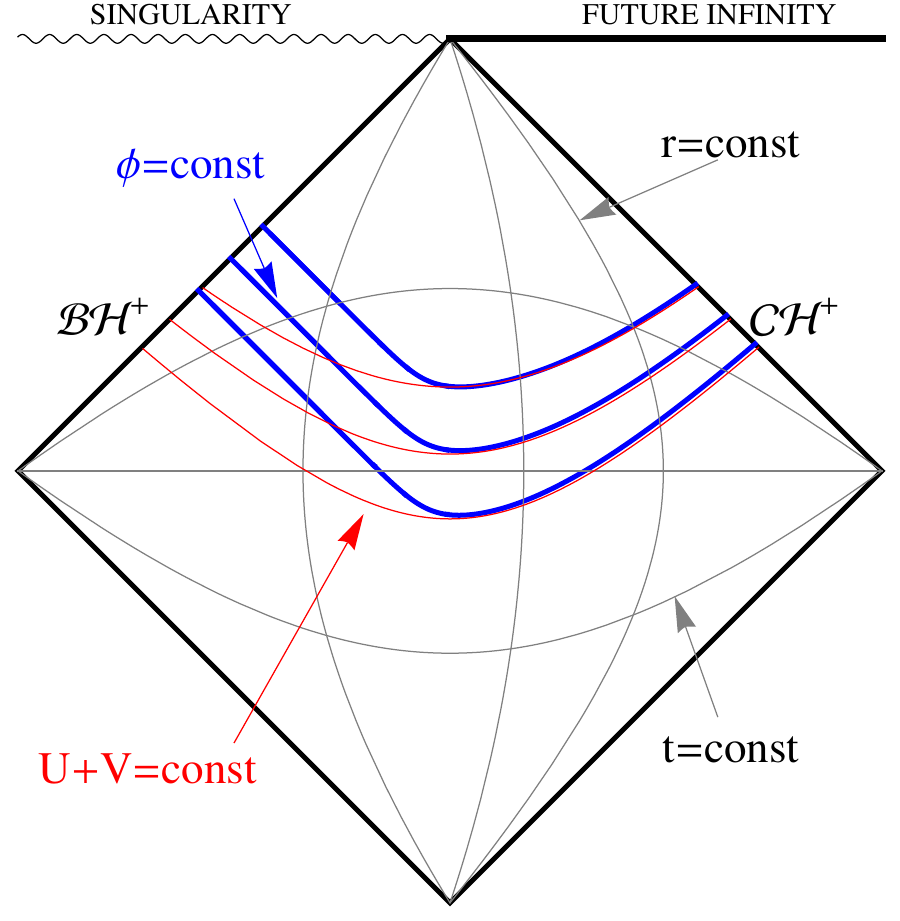}
\caption{Contours of constant $\phi$ (thick blue) within our 
Hubble volume. The Penrose diagram is obtained by 
compactifying along the diagonal directions $U$ and 
$V$ using the map $x \to (1+x)/(1-x)$. 
For comparison, contours of constant $t$ and $r$ are shown in grey, 
and a constant {\it cosmological} time contour ($U+V$) shown in red. 
The horizons are labelled, as are the singularity and future infinity.}
\label{fig:phicontours}
}

Figure \ref{fig:phicontours} shows that the rolling of the scalar lags
behind on the black hole event horizon, what is less clear is a slight lagging
on the cosmological event horizon. This is more clearly seen if we plot
$\phi$ as a function of $r^*$, 
as this makes the effect of the event horizons on the rolling of the
scalar clearer. Figure \ref{fig:phiofrstar} shows the profile
of the scalar field as a function of $r^*$ at differing values of
the `cosmological' time parameter $\eta=(U+V)/2$. The
lag due to the event horizon ($r^*\to-\infty$) is clearly shown
here, together with a slight lag (relative to $r^*\approx0$)
towards the cosmological event horizon, although the $\phi$ 
profile becomes flat at larger $r^*$, as indeed it should as
we expect to be close to the cosmological solution which
depends only on $U+V$. (For both figure \ref{fig:phicontours}
and \ref{fig:phiofrstar}, the parameter values $r_h=1$, $r_c=15$
were used.)
\FIGURE{
\includegraphics[scale=0.75]{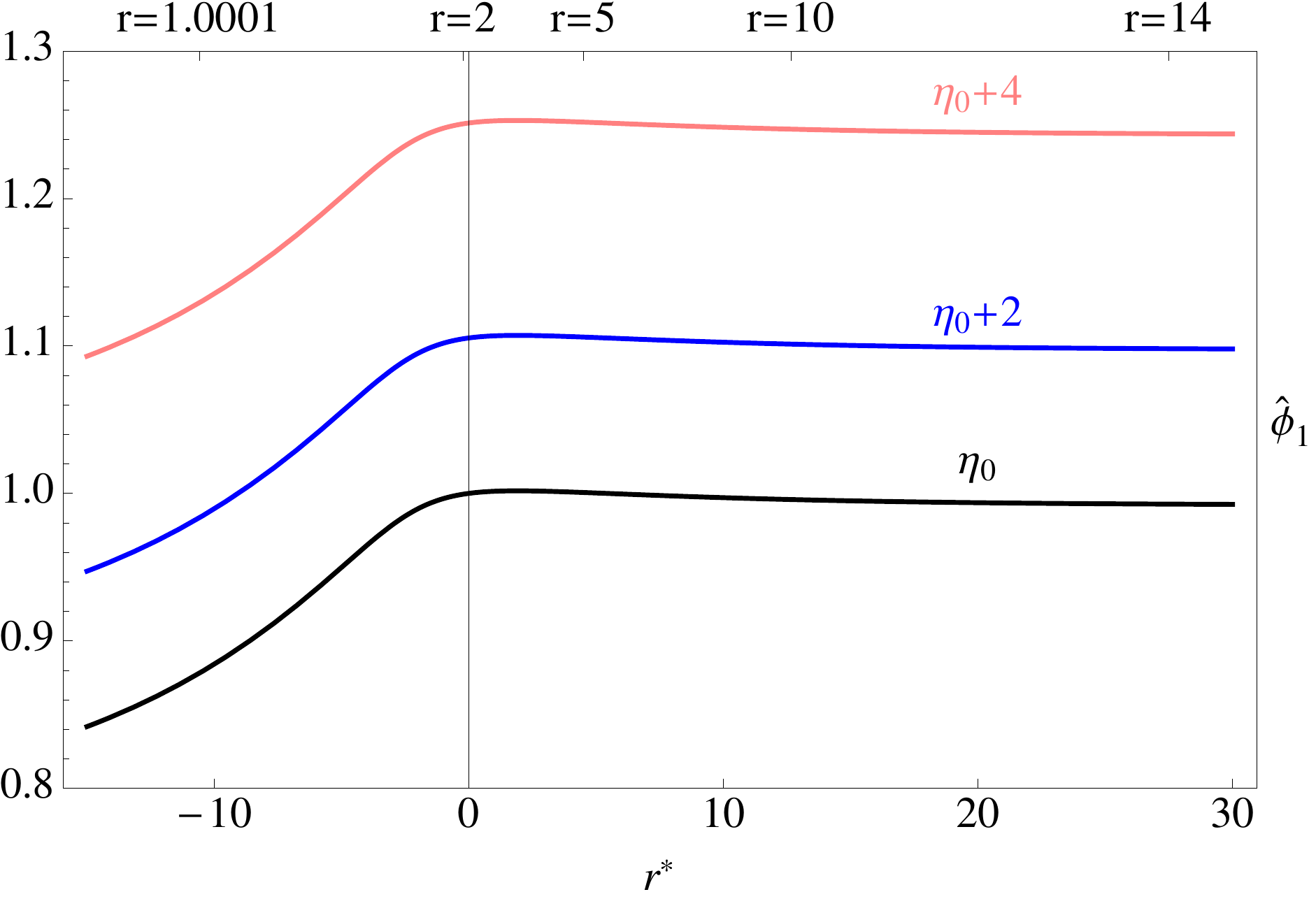}
\caption{The profile of the perturbation of the scalar field 
shown as a function of $r^*$. The field is illustrated at three 
different time steps, showing the field rolling to larger $\phi$ 
values. This figure was produced using the values $r_c=15$, 
$r_h=1$, $\eta_0=R_c\sim-8.3$, and for clarity the scalar 
field is normalised to its value at $\eta_0$ and
$r^*=0$, or alternately $t=-2R_c$, $r=2$: ${\hat \phi}_1 
= \phi_1/\phi_1(-2R_c,2)$ where, recall, $\phi_1=
\lambda t + \varphi$, given in \eqref{varphiexpression}.
(For reference, $\phi_1(-2R_c,2)\simeq-2.7$).
}
\label{fig:phiofrstar}
}

Finally, note that as $r_h\to0$, 
\be
\phi = \phi_0 + \sqrt{2\varepsilon} M_p\, \Bigl [ 
\frac{2R_c}{r_c} \ln \frac{U}{R_c} + \ln(r+r_c) \Bigr] 
\sim \phi_0 - \sqrt{2\varepsilon} M_p\, \ln |U+V|
\ee
i.e., the cosmological rolling scalar solution
to order ${\cal O}(\varepsilon^{1/2})$, (cf \eqref{phicost}).
On the other hand, near the black hole event horizon, 
\be
\phi \sim \phi_0 + \sqrt{2\varepsilon} M_p 2 \lambda R_h \ln v
\ee
in agreement with the linearized solutions of \cite{TJ,FK}.

\section{Back-reaction on the black hole geometry}

Clearly, since $\phi$ is regular from the black hole event horizon
out to the cosmological event horizon with regular derivatives, 
its energy momentum is finite in this region. We can therefore 
compute the back-reaction on the geometry to
get a consistent solution to order ${\cal O}(\varepsilon)$. 

We start by comparing the first integrals of \eqref{Vinteps}
\be
\beal
2\delta_2 &= 2\delta_1 + \int \frac{rV \phi_{1,V}^2}{R_cN} dV
+ \int \left [ \frac{rV\delta_{1,VV}}{2R_cN}  
+ \frac{r\delta_{1,V} }{2R_cN}
- \frac{(rN)'}{2N} \delta_{1,V} \right] dV + g(U)\\
&= \delta_1 + \int \frac{rV \phi_{1,V}^2}{R_cN} dV
+ \frac{rV}{2R_cN} \delta_{1,V} + g(U)
= \delta_1 + I_V
+ \frac{rV\delta_{1,V} }{2R_cN} + g
\eeal
\label{VVint}
\ee
and \eqref{Uinteps}
\be
\beal
2\delta_2 &= 2\delta_1 - \int \frac{rU \phi_{1,U}^2}{R_cN} dU
- \int \left [ \frac{rU\delta_{1,UU} }{2R_cN} 
+ \frac{r\delta_{1,U} }{2R_cN}
+ \frac{(rN)'}{2N} \delta_{1,U} \right] dU + f(V)\\
&= \delta_1 - \int \frac{rU \phi_{1,U}^2}{R_cN} dU
- \frac{rU}{2R_cN} \delta_{1,U} +f(V)
= \delta_1 +I_U
- \frac{rU\delta_{1,U} }{2R_cN} +f
\eeal
\label{UUint}
\ee
Where $f$ and $g$ are (for now) arbitrary integration functions, 
and can be thought of as the perturbation of $F'$ and $G'$.

Substituting for $\phi_1$ from \eqref{gendphi} shows that the
$\phi$-integrals in (\ref{VVint},\ref{UUint}) can be written as 
functions of $r$. For example
\be
\beal
\phi_{1,V} &= \frac{R_c}{V} \left ( \lambda + N\varphi'\right)
= \frac{R_c}{V} \left ( \lambda - H^2 r - \frac{C}{r^2} \right )\\
\Rightarrow \quad I_V&= \int \frac{rV \phi_{1,V}^2}{R_cN} dV 
= \int r \frac{dr}{dV} \left ( \lambda - H^2 r - \frac{C}{r^2} \right )^2
dV + \delta g(U)
\eeal
\ee
where the $\delta g(U)$ is added to acknowledge the fact that
an integral over $V$ can have an arbitrary integration factor
that is $U-$dependent, which may not be the same factor
as the $r$ integral. However, since our expressions in \eqref{VVint}
and\eqref{UUint} already contain integration functions, we
will now without loss of generality define
\be
\begin{aligned}
I_V &= \int \frac{rdr}{N^2} (H^2 r + \frac{C}{r^2}-\lambda)^2\\\
I_U &= \int \frac{rdr}{N^2} (H^2 r + \frac{C}{r^2}+\lambda)^2
\end{aligned}
\ee
and take it that the functions $f$ and $g$ are appropriately adjusted.

Next, consistency of (\ref{VVint},\ref{UUint}) requires
\be
U\delta_{1,U} +V\delta_{1,V} = 2R_c\frac{\partial~}{\partial t}
\delta_1 = \frac{2R_cN}{r} \left ( I_U-I_V+f-g\right)
\ee
which determines the general form of $\delta_1$ as
\be
\delta_1 = \frac{R_cN}{r} \left [ \ln(\frac{UV}{R_c^2}) (I_U-I_V)
+ \int \frac{2f}{V} dV - \int \frac{2g}{U} dU 
+ h(r) \right]
\label{delta1}
\ee
where $h(r)$ is an arbitrary integration function. Combining 
\eqref{VVint} and \eqref{UUint} then implicitly gives $\delta_2$:
\be
\delta_2 =  \frac{\delta_1}{2} + \frac{r}{4} \frac{\partial~}
{\partial r} \delta_1 + \frac{I_U+I_V}{4} + \frac{f+g}{4} \;.
\label{delta2form}
\ee
We now substitute these expressions into \eqref{BUVeps}, and
after some algebra, the only nonzero terms give
a second order ODE for $h(r)$:
\be
R_c \left [ rN^2 h' \right ] '= 12 H^2 r^2 \varphi 
+ (rN)' \left( I_U+I_V \right) \;.
\ee
It proves helpful to manipulate this equation using integration
by parts, and the fact that $12H^2r^2 = [rN^2(I_U'-I_V')]'/\lambda$,
to find an expression for h:
\be
h(r) = (I_U-I_V) \frac{\varphi}{\lambda R_c} + h_2(r) + h_3(r)
\ee
where
\be
\beal
h_2(r) &= \int \frac{I_U}{\lambda R_cN} \left ( \lambda + H^2 r + \frac{C}{r^2}\right)
+ \int \frac{I_V}{\lambda R_cN} \left ( \lambda - H^2 r - \frac{C}{r^2}\right)\\
h_3(r) &= \int \frac{1}{R_c rN^2} \int \frac{2r^2}{N}
\left ( \lambda + H^2 r + \frac{C}{r^2}\right)\left ( \lambda - H^2 r - \frac{C}{r^2}\right)
\eeal
\ee
Thus
\be
\delta_1 = \frac{R_cN}{r} \left [ \frac{\phi_1}{\lambda R_c} (I_U-I_V)
+ \int \frac{2f}{V} dV - \int \frac{2g}{U} dU 
+ h_2(r) + h_3(r) \right]
\label{delta1}
\ee

It is reasonably clear that $\delta_1$ is regular at both event 
horizons, provided $f$ and $g$ are no more divergent than
$V^{-1}$ or $U^{-1}$, however, we must examine regularity of
$\delta_2$, as we still need to determine the integration functions.
Inputting $\delta_1$ into \eqref{delta2form} gives
\be
\beal
\delta_2 = &\frac{\delta_1}{4} + \frac{f+g}{2}
+ \frac{R_cN'}{4} \left ( \int \frac{2f}{V} dV - \int \frac{2g}{U} dU 
\right)+ \frac{R_c}{4} \left[N(h_2+h_3)\right]' \\
&+ \frac{\phi_1}{4\lambda} \left [N(I_U-I_V)\right]' 
+ \frac{I_U}{4\lambda}\left ( \lambda - H^2 r-\frac{C}{r^2}\right)
+ \frac{I_V}{4\lambda}\left ( \lambda + H^2 r+\frac{C}{r^2}\right)
\eeal
\label{deltanu}
\ee
Clearly, the terms involving $I_U$, $I_V$ and $\delta_1$ are regular
from the definitions of $C$ and $\lambda$,
however, the residual pieces contain divergences, and we 
must choose $f$ and $g$ to regularise these. We will show this 
process in detail for the cosmological event horizon, the black 
hole event horizon follows the same steps. 

First we identify the potentially singular behaviour of the relevant 
functions
\be
\beal
I_U &= \frac{a_U}{r-r_c} + b_U \ln |r-r_c| + c_U + J_U(r)\\
I_V &= \frac{a_V}{r-r_h} + b_V \ln (r-r_h) + c_V + J_V(r)\\
N h_3 &= \frac{\alpha_c}{R_c} (r-r_c)\ln|r-r_c| 
+ \frac{\alpha_h}{R_h} (r-r_h) \ln(r-r_h) + N{\tilde h}_3
\eeal
\ee
where the constants can be inferred from the appendix,
$J_U={\cal O}(r-r_c)$, $J_V={\cal O}(r-r_h)$, and
$(N{\tilde h}_3)'$ is regular. Then the singular parts appearing in
\eqref{deltanu} are:
\be
\beal
&\frac{\phi_1}{\lambda} \left [N(I_U-I_V)\right]'\Big |_{\rm{sing}}
+ R_c\left [N(h_2+h_3)\right] ' \Big |_{\rm{sing}} 
=\frac{R_h}{R_c} b_V \left [ \ln(r-r_h)\right]^2 &\\
&
+\left [ \alpha_c + 2b_U+ 2 c_U + \frac{b_U\phi_1}{\lambda R_c}
+ 3 a_U H^2 \frac{r_h(r_h+r_c)}{(2r_c+r_h)}
\right]\ln |r-r_c|+b_U \left [ \ln|r-r_c|\right]^2
\qquad\\
&+ \frac{R_h}{R_c}
\left [ \alpha_h+2b_V+ 2 c_V + \frac{b_V\phi_1}{\lambda R_h}
+ 3a_V H^2 \frac{r_c(r_h+r_c)}{(2r_h+r_c)}\right]\ln (r-r_h)
&\eeal
\label{singh}
\ee
These can be cancelled by choosing $f=f_0 + f_1 \ln V$ 
and $g=g_0 + g_1 \ln U$, where the constants $f_i$ are chosen
to make $\delta_2$ regular at $r_c$, and the $g_i$ from
regularity at $r_h$. For example, as $r\to r_c$, 
\be
V \sim (2r_c+r_h)U \left ( \frac{r_h-r_c}{2r_c+r_h} 
\right)^{\frac{R_h}{R_c}}(r-r_c)
\ee
hence
\be
\beal
2f + \int\frac{2f}{V} dV \Bigg |_{\rm{sing}} &\sim
f_1 \left ( \ln V\right)^2 + (f_0+f_1) \ln V\\
& \sim f_1 \left [ \ln|r-r_c|\right]^2 + 2 f_1 \ln U \ln |r-r_c|\\
+ \Bigl [f_0+2f_1&+\frac{2f_1R_h}{R_c} \ln(r_c-r_h)
+ \frac{2f_1R_N}{R_c} \ln(r_c-r_N) \Bigr] \ln|r-r_c| 
\eeal
\ee
Comparing this with \eqref{singh}, and recalling that $\phi_1
\sim 2 R_C\lambda \ln(U/R_c)$, we see that
\bea
f_1 &=& -b_U\\
f_0&=& \alpha_c + 4b_U+ 2 c_U 
+\frac{2b_U}{R_c} \left [ R_h\ln(r_c-r_h)
+ R_N\ln(r_c-r_N)\right]
+ 3 a_U H^2 \frac{r_h(r_h+r_c)}{(2r_c+r_h)}
\nonumber
\eea
with similar expressions for the $g_i$.
The remaining, regular, parts of $\delta_1$ and $\delta_2$ are then
expressible in terms of regular dilogarithms and logarithms, but the full
expressions are rather lengthy and cumbersome.

Instead, by focussing on the event horizons, it is
easiest to obtain results of most physical interest. For the
cosmological event horizon, the $\{U,V\}$ coordinate system is
appropriate, with the CEH being at $V=0$, and parametrized by $U$. 
Since $r=$ const.\ along the horizons, \eqref{Uint} gives
\be
B = r_c^2 \left ( 1-8\varepsilon \lambda^2 R_c^2\ln |U| \right)
\simeq r_c^2 |U|^{-8\varepsilon \lambda^2 R_c^2}
\ee
Setting $r_h \to 0$ gives $4 \lambda^2 R_c^2 =1$, and hence $B 
\propto |U|^{-2\varepsilon}$, in complete agreement with the cosmological
event horizon area of the pure rolling scalar solution, \eqref{sccos}.

Of more interest however is the accretion of scalar field onto
the black hole. Here, using the Kruskal $\{u,v\}$ system and 
\eqref{Vint}, we get
\be
B = r_h^2 \left ( 1+8\varepsilon \lambda^2 R_h^2\ln v \right)
\simeq r_h^2 v^{8\varepsilon \lambda^2 R_h^2}
\label{hplusgrow}
\ee
In other words, the event horizon creeps out very slowly.
We can compare this with an order of magnitude estimate based
on  naive physical notions of mass and energy flow, \cite{TJ,FK}. 
The flow of energy into the black hole should be governed by 
the difference of the energy momentum tensor from being null, 
$T^0_0-T^r_r$, which is of order
${\dot\phi}^2 \sim W'^2/ H^2 \sim \varepsilon H^2/\kappa$.
Integrating this over the black hole event horizon gives
$\kappa\delta {\dot M} \sim r_h^2 \varepsilon H^2$, or 
using the relation between horizon radius and mass: 
$\delta A \sim r_h^3 \varepsilon H^2 \delta t $.
Of course, we should be careful of using a time coordinate
near the black hole event horizon, as 
$t \sim 2 r_h \ln v + \ln(r-r_h)$ is singular,
however, in the spirit of this heuristic argument, we can identify
$\delta t \sim 2r_h \delta (\ln v)$, which gives
$\delta A \sim A \varepsilon(\delta \ln v) r_h^2/r_c^2 $, 
in qualitative agreement with \eqref{hplusgrow}.

For an astrophysical black hole, this accretion rate is glacially
slow, and far outweighed by the local environment, in which
the accretion disc far outweighs local interstellar matter, let alone
this cosmologically coasting scalar. However, the fact that naive 
local estimates of the back-reaction of accretion of scalar matter 
in this set-up are fully backed up by this analytic calculation, valid
in the full region between the black hole and cosmological event
horizons, means we should have confidence in these physically 
motivated techniques.

To illustrate the effect of the rolling of the scalar, in figure 
\ref{fig:horizons} we show the effect on the event horizon
areas; for the purpose of illustration choosing $\varepsilon=0.1$,
and the rather artificial initial values of $r_c=2$, $r_h=1$.
Both horizons grow during the Hubble time, although the
cosmological event horizon has a larger relative growth of c.\ 22\%,
as opposed to approximately 13\% for the black hole event horizon.
\FIGURE{
\includegraphics[scale=0.8]{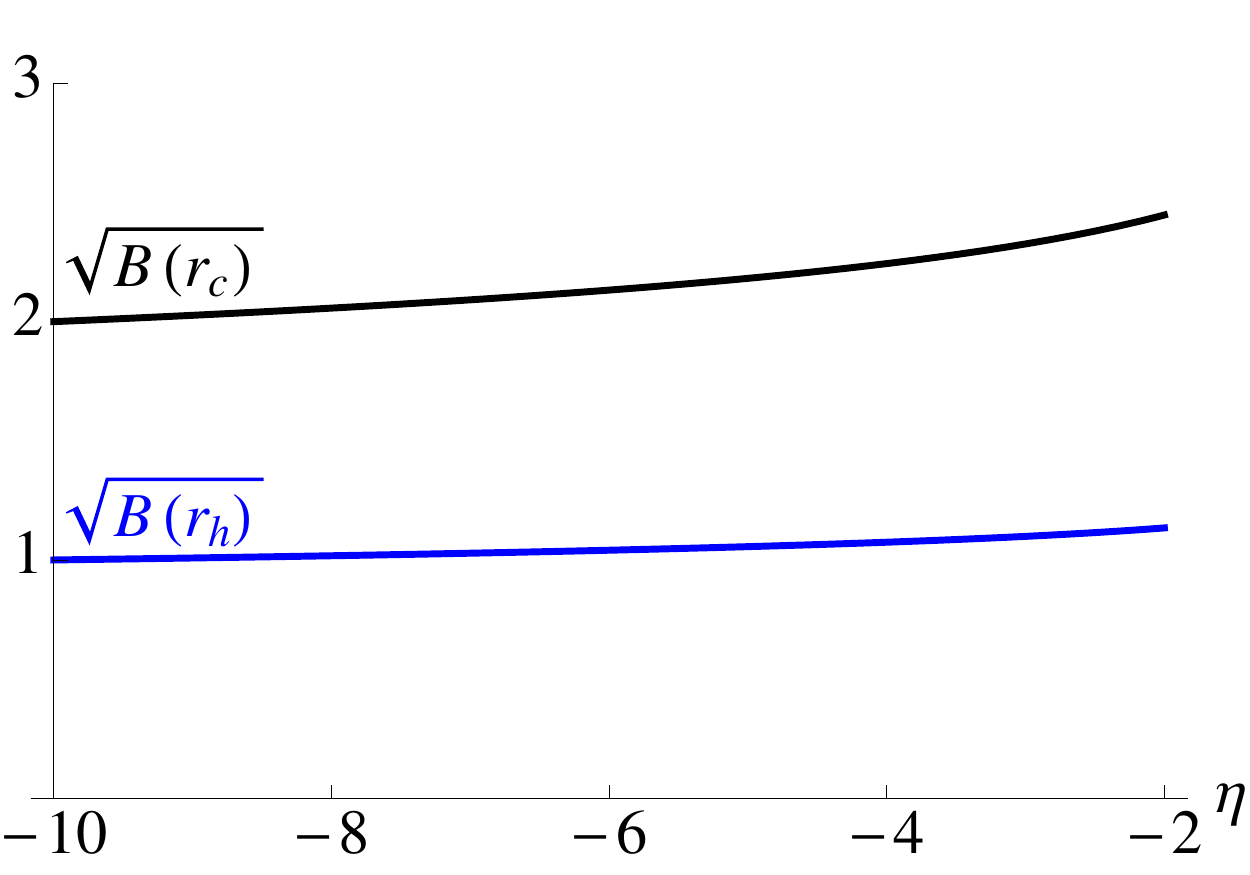}
\caption{A plot of the variation of the event horizon areas for the
initial choice of $r_c=2, r_h=1, \varepsilon=0.1$.}
\label{fig:horizons}
}

\section{Discussion}

In this paper, we have shown how to couple a slowly rolling scalar field
to a black hole. We first set up the most general metric describing
the physical set-up -- an SO(3) symmetric geometry with dependence
on both ``time'' and ``radial distance''. This geometry is most
naturally described in light-cone coordinates, which elucidate the 
integrable nature of the vacuum equations, and also lend themselves
to an accurate determination of the black hole event horizon.
Writing down the equations of motion reveals that if the potential is 
not too steep, the scalar field evolution will be slow, and the equations
of motion can be solved perturbatively in the dynamics of the scalar field.

It is worth emphasising that although we perform an expansion 
in the dynamics of the scalar field, the solutions we find are 
exact across the the full radial range from the black hole to 
cosmological event horizons, and are not in any sense perturbative
in the spatial (radial) coordinate. Not only do we correctly identify 
what the dynamical dependence is, we
are able to correctly identify the cosmological evolution (and hence
cosmological time) far from the black hole, as well as how the scalar
field drives this expansion, and how it is dragged by the black hole.
For example, we can apply these results to
black holes whose event horizon constitutes a significant fraction
of the Hubble volume. These black holes accrete at a similar rate
to the increase in area of the cosmological event horizon. 

It is interesting to compare the accretion rate of the black hole 
to the evaporation rate, to see whether black holes, or radiation,
will dominate the final state of the universe. The black hole 
evaporation rate is inversely proportional to the area of its 
event horizon: ${\dot M}_H \sim 10^{-4} \hbar/r_h^2$,
whereas the accretion rate is proportional to horizon area:
$G \delta M \sim  \varepsilon c H^2 r_h^2 \delta t$.
Thus in order for evaporation to dominate, the horizon
radius of the black hole would have to satisfy
\be
r_h \lesssim 0.1 \varepsilon^{-1/4} \left [
\frac{\hbar G}{ c H^2} \right] ^{1/4}
\sim 5 \times 10^{-6} \varepsilon^{-1/4} {\rm m}
\ee
Since $\varepsilon \sim 1+w$, where $w$ is the equation of state
for the dark energy, in order for astrophysical black holes to 
preferentially evaporate, we would require an equation of state fine tuned
to approximately $|1+w| \leq {\cal O} (10^{-36})$!

We should point out that while we have a time dependent
scalar field in a time dependent cosmological black hole background,
this should not be viewed as a violation of the no hair theorems. 
The solution corresponds to a rolling cosmological scalar field in
which there is a black hole, and the rolling of the scalar adjusts to
its presence. The scalar does roll in the vicinity
of the black hole, although it lags behind the cosmological
evolution, in the sense that the constant $\phi$
contours lie in front of the constant $\eta$ contours in figure
\ref{fig:phicontours}. However, the black hole does not have any
scalar charge -- there is no 1-parameter solution for $\phi$ in
the black hole background. The solution \eqref{gendphi} is
not the most general solution, there will be wave like fluctuations
around this background, but we have not been able to find a 
family of solutions with space-like dependence, which would
be a signature of a scalar charge on the black hole. Thus,
our results should be viewed as a way of reconciling the ``no hair''
intuition with more general time dependent situations.

Finally, to our knowledge, this is the first analytic procedure for 
finding a non-singular accreting black hole space-time from first 
principles (i.e.\ without making assumptions as to the form
of the metric or solution) that results from consideration of a physically 
realistic matter system with physically motivated symmetries and 
boundary conditions. The results apply to a general potential, 
and only require that the scalar field is slowly rolling.
As such, they represent a testing ground for investigation 
of black hole phenomena in the time dependent regime.

\acknowledgments
We would like to thank Anne Davis for helpful discussions.
RG is supported in part by STFC (Consolidated Grant ST/J000426/1),
in part by the Wolfson Foundation and Royal Society, and in part
by Perimeter Institute for Theoretical Physics. SC is supported by an 
EPSRC studentship. 
Research at Perimeter Institute is supported by the Government of
Canada through Industry Canada and by the Province of Ontario through the
Ministry of Economic Development and Innovation.

\appendix
\section{Appendix: Some useful identities}

Here we list some simple identities which are nonetheless
very useful in manipulation of expressions throughout the
paper.

First, we write the roots of the Schwarzschild potential as $r_i$ 
($r_N<0\leq r_h < r_c$) so that
\be
N(r)=-\frac{H^2}{r}(r-r_c)(r-r_h)(r-r_N)
\ee 
It is then simple to note the following identities:
\be
\beal
r_c+r_h+r_N &=0 \\
r_c^2 + r_h^2 + r_c r_h &= - r_cr_h - r_c r_N - r_hr_N = H^{-2}\\
r_c r_h r_N H^2 &= -2GM
\eeal
\ee

We have also defined
\be
R_i = \frac{1}{N'(r_i)} = \frac{r_i}{(1-3H^2r_i^2)}
\ee
which satisfy a similar identity to the $r_i$
\be
R_c+R_h+R_N=0
\ee

In addition, although we do not make use of the explicit forms of $I_U$ and
$I_V$, we note here their form for the constants used in determining
$f$ and $g$
\bea
I_U &=& - \frac{4 r_cR_c^2\lambda^2}{(r-r_c)} 
- \frac{r_N r_h^2 R_c^2\lambda^2}{(r-r_N)} 
+4 R_c^2 \lambda^2 H^2 ( 2r_c^3+r_h^3) \ln|r-r_c|
+ \frac{r_c^2r_h^2}{(r_c^2+r_h^2)^2}\ln r \nonumber \\
&&+ \frac{r_h}{r_c^2} \lambda^2 R_c^2 H^2 (4r_c^4+2r_c^3r_h
+3r_c^2r_h^2+5r_cr_h^3+r_h^4) \ln(r-r_N)\\
I_V &=& - \frac{4 r_hR_h^2\lambda^2}{(r-r_h)} 
- \frac{r_N r_c^2 R_h^2\lambda^2}{(r-r_N)} 
+4 R_h^2 \lambda^2 H^2 ( 2r_h^3+r_c^3) \ln(r-r_h)
+ \frac{r_c^2r_h^2}{(r_c^2+r_h^2)^2}\ln r \nonumber \\
&&+ \frac{r_c}{r_h^2} \lambda^2 R_h^2 H^2 (4r_h^4+2r_h^3r_c
+3r_c^2r_h^2+5r_hr_c^3+r_c^4) \ln(r-r_N)
\eea

\providecommand{\href}[2]{#2}\begingroup\raggedright\endgroup

\end{document}